\documentclass[article,nofootinbib,notitlepage]{revtex4-1}
\usepackage{graphicx}
\def\beq{\begin{equation}}
\def\eeq{\end{equation}}
\def\bea{\begin{eqnarray}}
\def\eea{\end{eqnarray}}

\def\gev{\text{ GeV}}

\begin{document}


\title{Dark matter annihilation with $s$-channel internal Higgsstrahlung}

\author{Jason Kumar}
\affiliation{Department of Physics \& Astronomy, University of Hawai'i, Honolulu, HI 96822, USA}

\author{Jiajun Liao}
\affiliation{Department of Physics \& Astronomy, University of Hawai'i, Honolulu, HI 96822, USA}

\author{Danny Marfatia}
\affiliation{Department of Physics \& Astronomy, University of Hawai'i, Honolulu, HI 96822, USA}



\begin{abstract}
We study the scenario of fermionic dark matter that annihilates to standard model fermions
through an $s$-channel axial vector mediator.  We point out that the well-known chirality suppression
of the annihilation cross section can be alleviated by $s$-channel internal Higgsstrahlung.  The shapes of
the cosmic ray spectra are identical to that of $t$-channel internal Higgsstrahlung
in the limit of a heavy mediating particle. Unlike the general case of $t$-channel
bremsstrahlung, $s$-channel Higgsstrahlung can be the dominant annihilation process even for
Dirac dark matter. Since the $s$-channel mediator can be a standard model singlet, collider searches for the mediator are easily circumvented.
\end{abstract}


\maketitle


\section{Introduction}

A key strategy in the search for dark matter (DM) is indirect detection: the search for cosmic rays
arising from dark matter annihilation in the cosmos.  But it is well-known that dark matter
annihilation to standard model (SM) fermion/anti-fermion pairs, a key signature, is suppressed if the dark matter is a real particle and flavor violation is minimal.  In this broad scenario, which
includes the constrained minimal supersymmetric standard model, one finds that the bremsstrahlung processes
$XX \rightarrow \bar f f Y$ ($Y = \gamma, Z, h$) can dominate over the $XX \rightarrow \bar f f$ annihilation process.
The study of such bremsstrahlung processes is central to indirect detection prospects in these
scenarios~\cite{Bergstrom:1989jr, Flores:1989ru, Barger:2009, Bell:2010ei, Ciafaloni:2011sa, Bell:2011if, Garny:2011cj, Barger:2011jg}.

Thus far, the focus of such studies has been on $t$-/$u$-channel annihilation, where the dominant contribution
to bremsstrahlung arises from the coupling of a SM boson to a new charged scalar.  These models yield predictable
spectra which, remarkably, depend largely on the choice of final state and are independent of the details of the
DM-SM interaction~\cite{Barger:2011jg}. However, the allowed parameter space is tightly constrained by LHC searches for the
charged mediator. 

In this work, we point out that the Higgsstrahlung processes $XX \rightarrow \bar f f h$, can dominate over $XX \rightarrow \bar f f$
in the case of $s$-channel annihilation, where the mediator is a SM gauge singlet.  This scenario is far less
constrained by LHC searches, but also yields predictions for cosmic ray spectra arising from dark matter
annihilation which can by utilized in indirect searches.
We focus on the case where the emitted boson is the SM Higgs boson, and the mediator is a new SM singlet boson.
But our results also apply to the scenario
in which the emitted boson is a new neutral scalar which may or may not decay to SM particles.  Regardless
of whether or not the scalar decays visibly, the associated $\bar f f$ spectrum will be unsuppressed, yielding an
enhancement in the cosmic ray signal over the $XX \rightarrow \bar f f$ annihilation process.

In Section II, we describe the general principles that underly the chirality suppression of $s$-wave dark matter
annihilation, and describe a model which lifts this suppression through $s$-channel Higgsstrahlung.  In Section III,
we compute the cross sections and spectra, and compare them to the previously studied case of $t$-channel Higgsstrahlung~\cite{Luo:2013bua}.
We conclude with a discussion of our results in
Section IV.

\section{General Principles}

The suppression of the $XX \rightarrow \bar f f$ process for the case of real dark matter and
minimal flavor violation (MFV) can be understood from general principles.  If the initial dark
matter state consists of two identical particles, then it must be invariant under charge conjugation.
Equivalently, the wave function must be totally symmetric (anti-symmetric) if the particle is a
boson (fermion).  Since the two-particle initial state is multiplied by $(-1)^{L+S}$ under charge
conjugation, an $s$-wave ($L=0$) initial state must have $S$ even; for either a spin-0 or spin-1/2
DM particle, this implies $S=0$, and thus $J=0$.  The final state then must also have
$J=0$, implying that the $\bar f f$ pair travel back-to-back with the same helicity.  The fermions must
arise from different SM Weyl spinors, and the matrix element must be proportional to the mixing of
the left- and right-handed spinors.  Such mixing violates SM flavor symmetries; if flavor violation
is minimal, then the $s$-wave annihilation matrix element must be suppressed by $m_f / m_X$.

This suppression is no longer required if the final state is $\bar f f Y$, where $Y$ is a SM
boson.  Previous work has focussed on the case in which
the boson is emitted from the virtual mediating particle, a process called
virtual internal bremsstrahlung (VIB). (Of course, if $Y$ is a gauge boson that couples to
$f$, $\bar f$, it will be emitted from an external line as well.)   This class of models is important because, if the boson cannot be emitted from an internal
line or the initial particles, then the process of boson emission is essentially the same as final state radiation, which is dominated by soft/collinear
emission for which the final state fermion propagator becomes nearly on-shell.  As a result, the soft/collinear contribution is suppressed
by a factor $m_f / m_X$, just as for the $XX \rightarrow \bar f f$ process~\cite{Bringmann:2007nk}.  Moreover, if the mediator and dark matter particle
are nearly degenerate in mass, then the VIB matrix element is enhanced in the region of phase space where one final state
fermion is soft, and the propagator of the mediator is nearly on-shell.

If the boson emitted through VIB is a photon, then the mediator is charged under
$U(1)_{em}$, implying that it must be exchanged in the $t$- or $u$-channel.
But if the emitted boson is a scalar, then it may be emitted from a SM singlet particle.  In this case, VIB can occur
even if dark matter annihilates in the $s$-channel through a SM singlet mediator.

\subsection{Model}

We consider the case where the mediator is a heavy real spin-1 particle, $B_\mu$, which couples to
fermion dark matter ($X$) and SM matter through the following Lagrangian:
\bea
{\cal L}_{int} &=& {\lambda_X \over 2} \bar X \gamma^\mu \gamma^5 X B_\mu
+ \lambda_f \bar f \gamma^\mu (\sin \theta + \cos \theta \gamma^5) f B_\mu
+ {\lambda_h \over 4} H^2 B^\mu B_\mu ,
\eea
where $f$ is a SM fermion, $H = \langle H \rangle + h$ is the Higgs field and $\langle H \rangle = v_{EW} \sim 246~\gev$.  This interaction
structure (fermion dark matter and a spin-1 mediator which couples to an axial vector dark matter current
and a vector and/or axial vector SM current) is the only one that is suitable for our purpose.
Higgsstrahlung is relevant only if the DM-mediator interaction term has an unsuppressed matrix
element with an $s$-wave initial state, and if  the SM-mediator interaction is necessarily suppressed for the kinematics of
a two-particle final state when the outgoing SM particles are relativistic.
The appropriate suppression of the SM-mediator interaction for a two-body final state only occurs for the time-like component
of a spin-1 mediator, coupling to either a vector or axial-vector SM fermion current~\cite{Kumar:2013iva} (in the axial vector
current case, the interaction is suppressed by $m_f$, and in the vector current case it vanishes identically).
The mediator must then couple to a vector or axial vector dark matter current, such that only the time-like component
of the dark matter current has
an unsuppressed matrix element with an $s$-wave initial state.  This requirement is only satisfied if the dark matter is
spin-1/2 and couples to the mediator through an axial vector interaction~\cite{Kumar:2013iva}.
Note, if dark matter is spin-1 and couples to the
mediator through a vector interaction ($X^\nu \partial_\nu X^\mu B_\mu $), then the time-like component of the DM current does indeed have a non-trivial matrix
element for an $s$-wave initial state, but this matrix element vanishes in the non-relativistic limit
because it involves time-like polarizations of the DM particles~\cite{Kumar:2013iva}.

The shapes of the energy spectra for the process $XX \rightarrow \bar f f h$, summed over final state spins, are independent of $\theta$ in the
$m_f / m_X \rightarrow 0$ limit.  In this limit, $\theta$ only determines the relative branching fraction to final
states with left-handed and right-handed $f$.  For simplicity, we set $\theta=0$.

It is interesting to also note that in this scenario, the dark matter fermion $X$ can be either Dirac or Majorana, while
still exhibiting chirality suppression of the $\bar XX \rightarrow \bar f f$ cross section, which is lifted by $s$-channel
Higgsstrahlung.  This differs from the case of $t$-channel Higgsstrahlung, for the which the dark matter must be Majorana.
This is because if dark matter interacts with SM matter through the $t$- or $u$-channel, one must use a Fierz
transformation to construct the dark matter current which acts on the initial state.  Generically, one gets a linear
combination of all possible DM currents, including those which have a non-trivial matrix element with an $L=0$, $S=1$, $J=1$
initial state.  If the initial state is $J=1$, then the final state is $J=1$ as well, and the chirality suppression in the $m_f / m_X \rightarrow 0$ limit no longer applies.  In the $t$-channel case, it is
thus necessary to assume that dark matter is Majorana in order to eliminate the $J=1$ contribution.  For the $s$-channel case,
no such assumption is necessary because the choice of interaction Lagrangian picks out a particular dark matter current that
couples to the $s$-channel mediator; if the DM current is axial vector, then it has a trivial matrix element with the $L=0$,
$S=1$, $J=1$ state.  Such an interaction Lagrangian naturally arises for Dirac dark matter if the mediator is an
axial vector, and if the DM-mediator interaction respects $C$ and $P$.

\section{Cross Sections and Spectra}

The $XX \rightarrow \bar f f$ cross section is given by
\bea
v_{rel}\sigma(XX \rightarrow \bar f f)
&=& {\lambda_X^2 \lambda_f^2 N_c \over 2\pi} {m_f^2 \over \left(m_B^2 - 4m_X^2 \right)^2}\,,
\eea
where $N_c$ is the color factor associated with $f$, and $v_{rel}$ is the relative velocity of the initial state particles.  As expected, it vanishes in the limit
$m_f / m_X \rightarrow 0$.

The amplitude for the process, $X(k_1)X(k_2)\rightarrow f(p_1)\bar{f}(p_2)h(k)$, can be written as
\bea
i{\cal M}=\lambda_X\lambda_f\left(i\lambda_h v_{EW}\right) \frac{\left[\bar{v}(k_2)\gamma^\mu\gamma^5u(k_1)\right]\left[\bar{u}(p_1)\gamma_\mu\gamma^5v(p_2)\right]}{\left[(k_1+k_2)^2-m_B^2\right]\left[(p_1+p_2)^2-m_B^2\right]}\,.
\eea
The differential cross section in the limit $m_f\rightarrow 0$ is
\bea
v_{rel}\frac{d\sigma}{dx_1dx_2}=\frac{\lambda_X^2\lambda_f^2\lambda_h^2v_{EW}^2 N_c}{32\pi^3m_X^4}\frac{4x_1x_2-(4+r_h-4x_h)}{(4-r_B)^2(4+r_h-4x_h-r_B)^2}\,,
\label{eq:dsdx1dx2}
\eea
where $r_B \equiv m_B^2/m_X^2$, $r_h \equiv m_h^2/m_X^2$, and similar to the notation in Ref.~\cite{Barger:2011jg},
we define $x_1 \equiv E_f/m_X$, $x_2 \equiv E_{\bar{f}}/m_X$ and $x_h \equiv E_h/m_X$, so that in the static center of mass frame $x_1+x_2+x_h=2$.

The energy distribution of $f$ can be obtained by integrating over $x_2$ from $1-x_1-r_h/4$ to $1-r_h/(4(1-x_1))$~\cite{Fukushima:2012sp}, yielding
\bea
v_{rel}\frac{d\sigma}{dx_1}=\frac{\lambda_X^2\lambda_f^2\lambda_h^2v_{EW}^2 N_c}{128\pi^3m_X^4(4-r_B)^2}
\left[(1-x_1)\ln\left(\frac{r_B}{r_B-x_1(4-r_h-4x_1)/(1-x_1)}\right)-\frac{x_1(4-r_h-4x_1)}{r_B}\right]\,.
\label{fermion}
\eea
In the large $r_B$ limit, we have
\bea
v_{rel}\left.\frac{d\sigma}{dx_1}\right| _{r_B\rightarrow\infty}=\frac{\lambda_X^2\lambda_f^2\lambda_h^2v_{EW}^2 N_c x_1^2(4-r_h-4x_1)^2}{256\pi^3m_X^4r_B^4(1-x_1)}\,.
\eea

The Higgs spectrum can be obtained by integrating over $x_1\in[x_1^-,x_1^+]$ with $x_1^\pm=\frac{1}{2}(2-x_h\pm\sqrt{x_h^2-r_h})$, yielding
\bea
v_{rel}\frac{d\sigma}{dx_h}=\frac{\lambda_X^2\lambda_f^2\lambda_h^2v_{EW}^2 N_c}{48\pi^3m_X^4}\frac{(x_h^2-r_h)^{3/2}}{(4-r_B)^2(4+r_h-4x_h-r_B)^2}\,.
\eea
We do not consider the regime $2m_X\geq m_B+m_h$, as in this case  the $s$-channel annihilation cross section
would be dominated by the on-shell $2\rightarrow 2$ process, $XX\rightarrow Bh$.
If $2m_X > m_B$, then it is possible to produce an on-shell $B$ and an off-shell $h$, which in turn couples to SM particles.
  Unless $\lambda_f$ is very small, $B$ has a larger decay width than the Higgs, which implies that on-shell $h$ production dominates
over on-shell $B$ production.
Thus, in the entire mass range $m_h <  2m_X < m_B + m_h$, we are justified in considering only final states with
an on-shell $h$.

The $t$-channel differential cross section can be written as
\bea
v_{rel}\frac{d\sigma}{dx_1dx_2}=\frac{y_{DM}^4\lambda_h^2v_{EW}^2}{256\pi^3m_X^4}\frac{4x_1x_2-(4+r_h-4x_h)}{(1-2x_1-r_B)^2(1-2x_2-r_B)^2}\,,
\label{eq:tdsdx1dx2}
\eea
where $y_{DM}$ is the coupling between $X$, the mediator, and SM matter. By a slight abuse of notation, we denote the $t$-channel scalar mediator by $B$. We successfully reproduced the primary $t$-channel Higgsstrahlung spectra of Ref.~\cite{Luo:2013bua} (but not the secondary spectra, as we comment on below).

\begin{figure}
\centering
\centering
\includegraphics[width=0.49\textwidth]{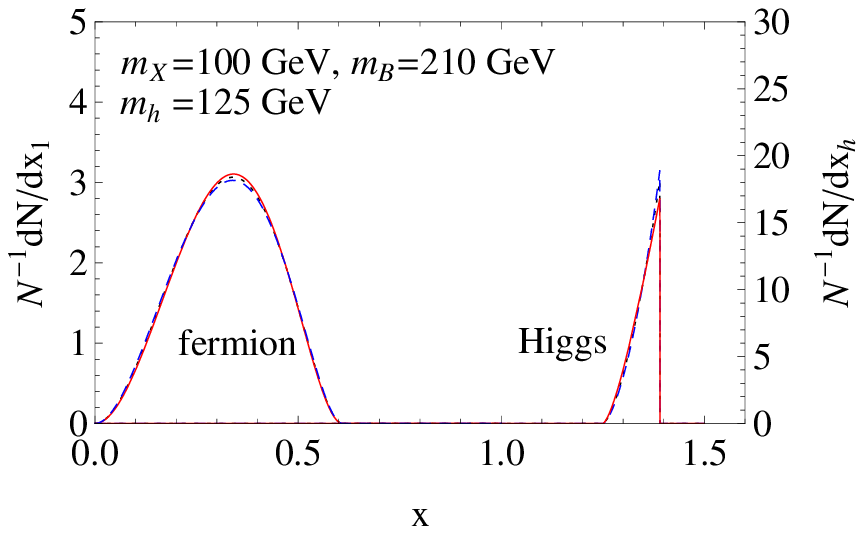}
\includegraphics[width=0.49\textwidth]{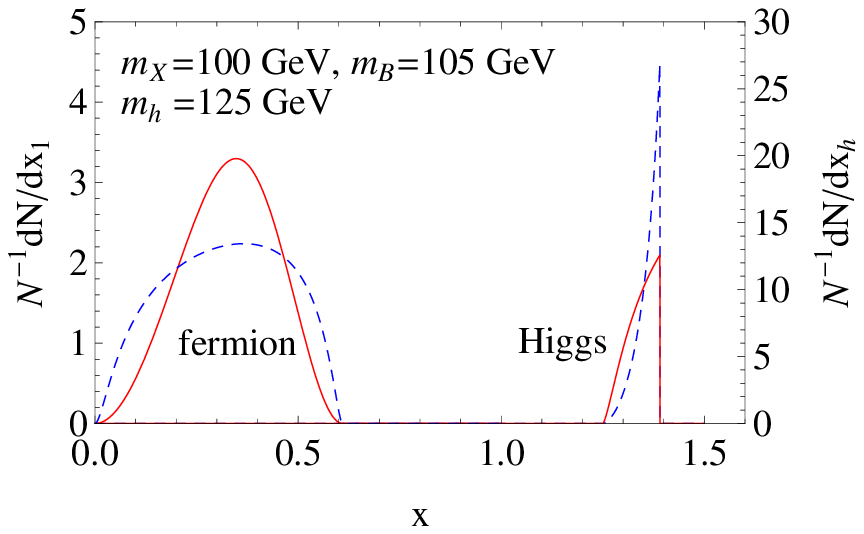}
\caption{Normalized fermion and Higgs spectra.
The red solid (blue dashed) 
curves correspond to $s$-channel ($t$-channel) Higgsstrahlung.  The black dotted curves in the left panel correspond to the $r_B \rightarrow \infty$ limit.  }
\label{fg:fermion}
\end{figure}

\begin{figure}
\centering
\centering
\includegraphics[width=0.5\textwidth]{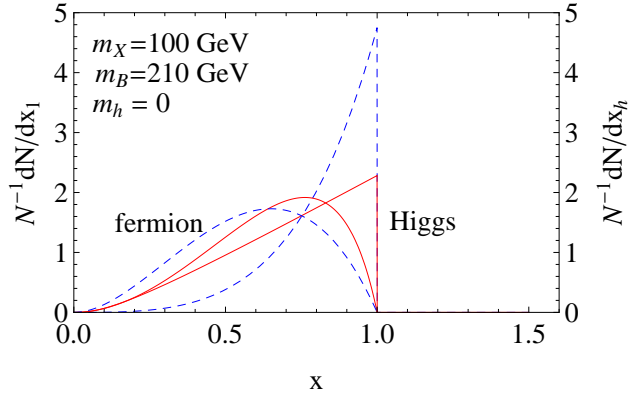}
\caption{Similar to Fig.~\ref{fg:fermion}, except for a new neutral Higgs with $r_h=0$.
}
\label{fg:higgs}
\end{figure}

Comparing Eqs.~(\ref{eq:dsdx1dx2}) and~(\ref{eq:tdsdx1dx2}), we see that the main difference between the $s$-channel and $t$-channel annihilation is the propagator. In the limit of a heavy mediator, $s$-channel and $t$-channel Higgsstrahlung yield the same normalized primary spectra, making them impossible to distinguish; see the left panel of Fig.~\ref{fg:fermion}.
From Fig.~\ref{fg:higgs}, we see that for $m_h \ll m_X$, $m_B$, the $s$-channel and $t$-channel spectra
are distinguishable because $r_B$ no longer dominates the denominators of Eqs.~(\ref{eq:dsdx1dx2}) and~(\ref{eq:tdsdx1dx2}).

The similarity of the spectra arising from $s$-channel and $t$-channel Higgsstrahlung in the heavy mediator limit is easily
understood.  In the heavy mediator limit, the mediator can be integrated out and the matrix element for the $XX \rightarrow \bar f f h$
annihilation process can be derived from an effective contact operator.  Since we and Ref.~\cite{Luo:2013bua} have assumed MFV
and taken the $m_f / m_X \rightarrow 0$ limit, the operators relevant for either $s$- or $t$-channel bremsstrahlung
cannot mix left-handed and right-handed $f$ Weyl spinors.  Moreover, because there is no mixing of SM Weyl spinors, and because
$X$ is a SM singlet, $SU(2)_L$ gauge-invariance requires an explicit insertion of a Higgs vev, $v_{EW}$.  The
relevant contact operator must therefore
be at least dimension 8.
There are two dimension 8 contact operators which satisfy these constraints and have non-trivial matrix elements
with an $L=0$ dark matter initial state:
\bea
{\cal O}_{AA} &=& {1 \over 2 \Lambda^4} (\bar X \gamma^\mu \gamma^5 X)(\bar f \gamma_\mu \gamma^5 f) H^2
\rightarrow {v_{EW} \over \Lambda^4} (\bar X \gamma^\mu \gamma^5 X)(\bar f \gamma_\mu \gamma^5 f) h\,,
\nonumber\\
{\cal O}_{AV} &=& {1 \over 2 \Lambda^4} (\bar X \gamma^\mu \gamma^5 X)(\bar f \gamma_\mu  f) H^2
\rightarrow {v_{EW} \over \Lambda^4} (\bar X \gamma^\mu \gamma^5 X)(\bar f \gamma_\mu  f) h\,.
\eea
Note that in the heavy mediator limit, one expects $\Lambda \propto m_B$, implying that the Higgsstrahlung
cross section scales as $r_B^{-4}$, as expected.
In the heavy mediator limit, $s$- and $t$-channel higgsstrahlung are produced by different linear combinations
of ${\cal O}_{AA}$ and ${\cal O}_{AV}$.  But in the $m_f / m_X \rightarrow 0$ limit, these operators produce
the same energy spectra.  They differ only in the relative sign  of the matrix element for coupling to
left-handed and right-handed $f$, but this sign is unobservable in the chiral limit.
Although this argument is only valid in the contact-interaction limit, we see that for $r_B > 4$ the normalized spectra are already
quite similar.

Finally, the total cross section can be expressed as
\bea
v_{rel}\sigma(XX \rightarrow \bar f f h) &=& \frac{\lambda_X^2\lambda_f^2\lambda_h^2v_{EW}^2 N_c}{4096\pi^3m_X^4(4-r_B)^2}\left\{(\Lambda+8r_h)\ln\frac{2}{\sqrt{r_h}}\right.
\\\nonumber
& &\left. -\frac{4-r_h}{6r_B}\left[6r_B^2+2(4-r_h)^2-9r_B(4+r_h)\right] \right.
\\\nonumber
& &\left. +(4-r_B+r_h)\sqrt{\Lambda}\ln\left[\frac{4r_B\sqrt{r_h}}{r_B(4+r_h)-(4-r_h)(4-r_h+\sqrt{\Lambda})}\right]\right\}\,,
\eea
where $\Lambda \equiv 16+r_B^2+r_h^2-8r_B-8r_h-2r_Br_h$.  As expected, there is a resonant enhancement as $r_B \rightarrow 4$.

In the large $r_B$ limit, the total cross section becomes
\bea
v_{rel}\sigma(XX \rightarrow \bar f f h)\mid_{r_B\rightarrow\infty} &=& \frac{\lambda_X^2\lambda_f^2\lambda_h^2v_{EW}^2 N_c}{192\pi^3m_X^4r_B^4}\left[1-2r_h +\frac{r_h^3}{8}-\frac{r_h^4}{256}+\frac{3}{2}r_h^2\ln\frac{2}{\sqrt{r_h}}\right]\,.
\eea
It is easy to verify that the above equation has the same form as Eq.~(A.3) in Ref.~\cite{Luo:2013bua}, up to a normalization factor and a change of variable.

\subsection{Secondary Spectra}

In our model, the injected cosmic ray spectrum arises both from the direct
injection of $\bar f f$ pairs, and from the decay products of the Higgs boson. 
More generally, the mediator could couple to any real scalar
$\phi$ via \mbox{${\cal L}_\phi = (\lambda_\phi /2) v_{EW} \phi B^\mu B_\mu$},
where the factor of $v_{EW} \sim 246~\gev$ is included as a convenient energy scale for the coupling.
The primary $\bar f f$ spectrum would be as in Eq.~(\ref{fermion}), with the emitted scalar boson mass a free parameter.  The part of the cosmic ray spectrum arising from scalar
decay would now depend on the branching fractions for $\phi$ to decay to various SM final states,
and could be absent entirely if $\phi$ decayed invisibly.
The features of these total spectra thus depend in detail on the choice of $f$, as well as on the
visible decays of the scalar.  

We use the {\it cookbook} of Ref.~\cite{pppc} to obtain the spectra of stable particles at the source (including decays, showering and hadronization) for a few special cases in which we assume the mediator couples equally to first generation leptons, and does not couple to other SM matter fields. From Fig.~\ref{spectra}, we see that in each case, including the $m_X=100$~GeV and $m_B=105$~GeV case, the resultant $s$-channel and $t$-channel spectra are similar for the SM Higgs.{\footnote{Note that we could not reproduce the $t$-channel distributions of the positron and neutrino in Ref.~\cite{Luo:2013bua}. As a check that we are using the ingredients
of Ref.~\cite{pppc} correctly, we reproduced the electroweak bremsstrahlung spectra of Ref.~\cite{Bell:2010ei}.}

As an example of a new real scalar, we consider a Higgs-like boson with a mass that lies between $2m_\mu$ and $2m_\pi$, as may occur in models with Higgs portals. Such a boson decays dominantly to muons. As can be seen from Fig.~\ref{spec}, the Higgsstrahlung signatures can be very different from that for the SM Higgs. 

\begin{figure}
\centering
\centering
\includegraphics[width=0.49\textwidth]{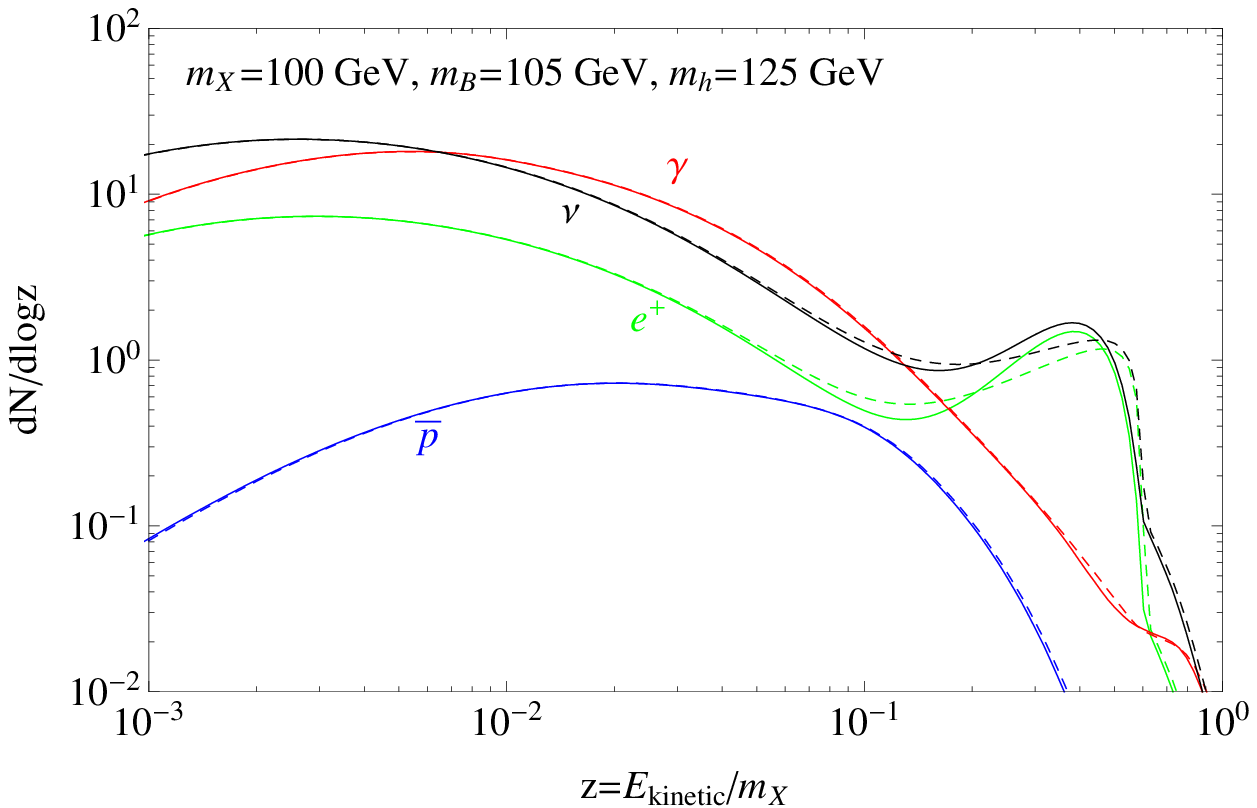}
\includegraphics[width=0.49\textwidth]{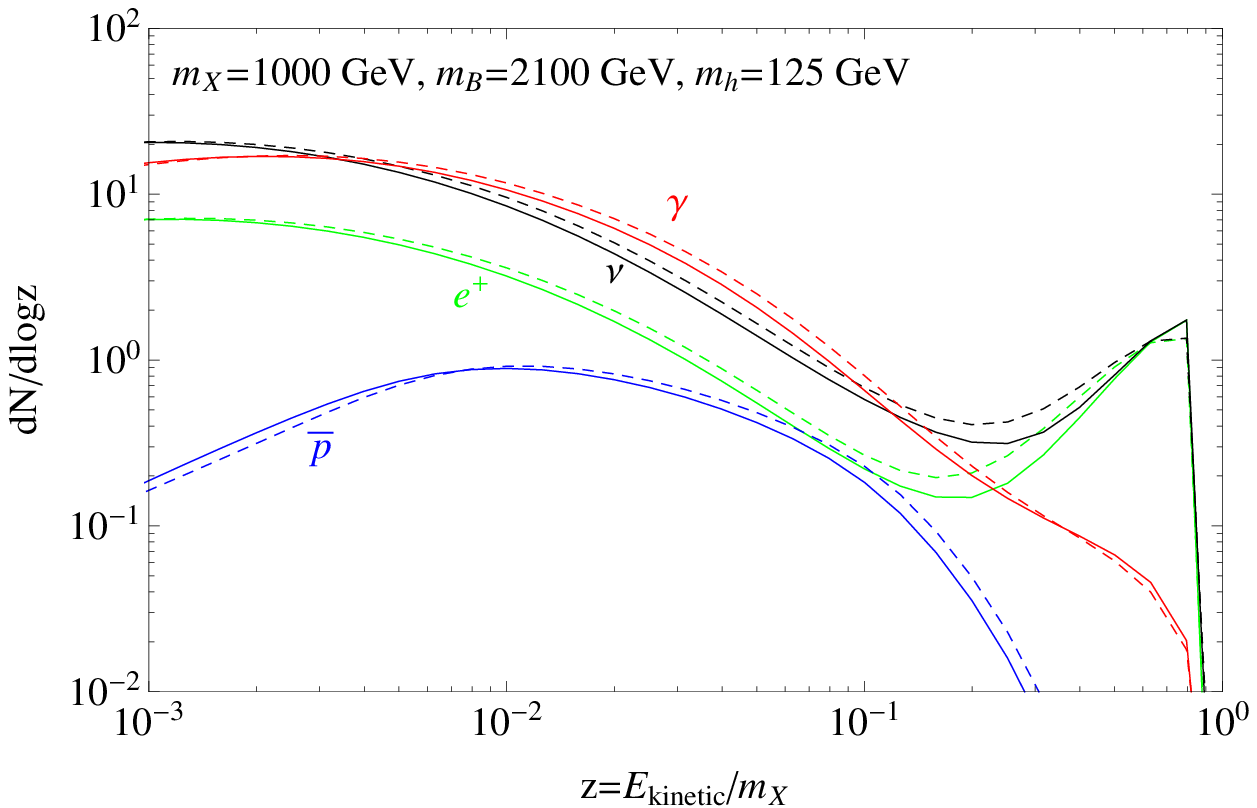}
\caption{Spectra of stable particles at the source (normalized to the multiplicities per annihilation) assuming that the mediator couples equally to first generation leptons, and does not couple to other SM matter fields. The combined $\nu_e + \nu_\mu + \nu_\tau$ spectrum is denoted by $\nu$.
The solid (dashed) curves correspond to $s$-channel ($t$-channel) Higgsstrahlung for the SM Higgs boson. }
\label{spectra}
\end{figure}

\begin{figure}
\centering
\centering
\includegraphics[width=0.49\textwidth]{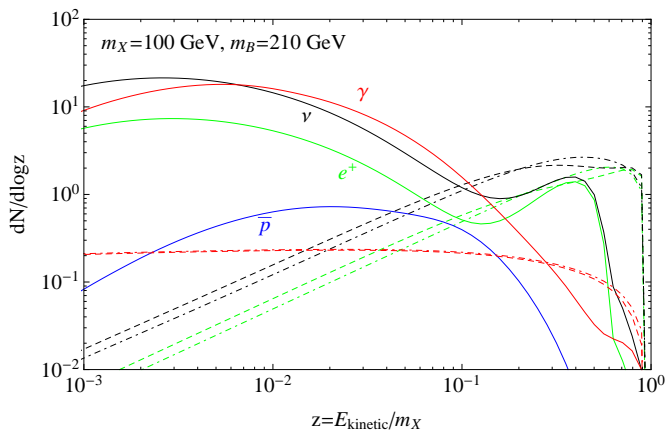}
\caption{The solid curves correspond to both $s$-channel and $t$-channel Higgsstrahlung (since they are indistinguishable) for the SM Higgs boson with $m_h=125$~GeV.  The dashed (dot-dashed) curves correspond to the $s$-channel ($t$-channel) processes with a light Higgs-like boson of mass 250~MeV that decays dominantly to muons. No antiprotons are produced by this boson. The mediator has the same couplings as in Fig.~\ref{spectra}.}
\label{spec}
\end{figure}

\section{Conclusions}

We calculated the differential cross section for the $s$-channel Higgsstrahlung process
$XX \rightarrow \bar f f h$.  This scenario arises when a spin-1 mediating particle has
vector or axial vector couplings to a SM fermion $f$, axial vector coupling to a fermion dark matter particle $X$,
and a coupling to the Higgs boson.
The spectra reduce to the previously known $t$-channel Higgsstrahlung spectra in the 
contact-interaction
limit.  But there are differences in the viability of these scenarios, given data from the LHC.
$t$-channel Higgsstrahlung necessarily involves a electroweak and/or QCD charged mediator, and there are tight
constraints on the masses of such particles from current LHC data.  Since an $s$-channel mediator may be a
SM singlet, it can evade such bounds, opening up new regions of parameter space where Higgsstrahlung is relevant
to dark matter annihilation.

Unlike the case of $t$-channel annihilation, $s$-channel annihilation can receive a chirality suppression which
is lifted by Higgsstrahlung even if the dark matter is a Dirac fermion.  This provides an interesting correlation
between cosmic ray signatures of dark matter annihilation and the properties of dark matter, assuming that dark matter
is stable.  In particular,
in the case of $t$-channel annihilation, the dominance of internal bremsstrahlung processes over
chirality-suppressed $XX \rightarrow \bar f f$ annihilation processes would imply that dark matter must be a
real particle.  Since a real particle cannot be charged under an exact continuous symmetry, this would imply that
dark matter was stabilized by a discrete symmetry.  But if dark matter annihilates through the $s$-channel, then it
may be stabilized by a continuous symmetry and still exhibit a chirality-suppressed $XX \rightarrow \bar f f$ annihilation
cross section; the chirality suppression can then be lifted by Higgsstrahlung.

Although we have focused on the Higgsstrahlung process $XX \rightarrow \bar f f h$, the Higgs boson may be
replaced by any new scalar particle $\phi$ without altering the form of the primary fermion spectrum.  In this
case, both $m_X$ and $m_\phi$ may be well below the electroweak scale.  The annihilation of low mass dark matter
to either $b$-quarks or $\tau$-leptons has been considered as a possible source of the excess in GeV-scale photons
observed from the Galactic Center (GC), and detailed fits of the observed photon spectrum from the GC to the
spectra expected from the processes $XX \rightarrow \bar b b, \bar \tau \tau$ have been performed~\cite{Daylan:2014rsa}.  But these processes
are relevant only if $s$-wave dark matter annihilation to fermions is not very chirality-suppressed; if it is suppressed,
then scalar bremsstrahlung processes could dominate.  The softening
of the primary fermion injection spectrum arising from the process $XX \rightarrow \bar f f \phi$ would change the
spectrum of photons produced at the GC.  It would be interesting to reconsider the GC excess in this light.

\acknowledgments

JK is supported in part by NSF CAREER grant PHY-1250573.  JL and DM are supported in part by
DOE grant DE-SC0010504.





\end{document}